\documentclass{article}

 \pdfoutput=1
 \usepackage{hyperref}
 \hypersetup{
   pdfinfo={
     Title={A Multi-Phase Gammatone Filterbank for Speech Separation via TasNet},
     Author={David Ditter and Timo Gerkmann},
     Subject={},
     Keywords={Speech Separation, Auditory Filterbank, End-To-End Learning, TasNet}
   }
 }
 \usepackage{pdfpages}

\usepackage{spconf,amsmath,graphicx}
\usepackage[utf8]{inputenc}
\usepackage{lipsum}
\usepackage{siunitx}
\usepackage[nolist]{acronym}
\usepackage[keeplastbox]{flushend}

\usepackage{tikz}
\usetikzlibrary{calc, matrix}
\usepackage{pgfplots}

\pgfplotsset{compat=1.8}
\usepackage{pgfplotstable}
\usepgfplotslibrary{groupplots}

\usepackage{amssymb}%
\usepackage{pifont}%

 \title{A Multi-Phase Gammatone Filterbank for Speech Separation via TasNet}
\name{David Ditter and Timo Gerkmann}
\address{Signal Processing (SP), Universität Hamburg, Germany\\
david.ditter@uni-hamburg.de, timo.gerkmann@uni-hamburg.de
}
\begin{document}
\begin{acronym}
\acro{agtf}[A-GTF]{auditory gammatone filterbank}
\acro{BLSTM}[BLSTM]{Bidirectional LSTM}
\acro{convtasnet}[Conv-TasNet]{convolutional time domain audio separation network}
\acro{erb}[ERB]{equivalent rectangular bandwidth}
\acro{fir}[FIR]{finite impulse response}
\acro{iir}[IIR]{infinite impulse response}
\acro{LPC}[LPC]{Linear Predictive Coding}
\acro{LSTM}[LSTM]{Long Short-Term Memory}
\acro{mpgtf}[MP-GTF]{multi-phase gammatone filterbank}
\acro{relu}[ReLU]{Rectified Linear Unit}
\acro{SDR}[SDR]{Source-to-Distortion Ratio}
\acro{sisnr}[SI-SNR]{scale-invariant source-to-noise ratio}
\acro{STFT}[STFT]{Short-Time Fourier Transform}
\acro{tasnet}[TasNet]{TasNet}
\acro{PIT}[PIT]{Permutation Invariant Training}
\acro{VTL}[VTL]{Vocal Tract Length}

\end{acronym}

\ninept
\maketitle
\begin{abstract}
In this work, we investigate if the learned encoder of the end-to-end \ac{convtasnet} is the key to its recent success, or if the encoder can just as well be replaced by a deterministic hand-crafted filterbank. Motivated by the resemblance of the trained encoder of \ac{convtasnet} to auditory filterbanks, we propose to employ a deterministic gammatone filterbank. In contrast to a common gammatone filterbank, our filters are restricted to 2 ms length to allow for low-latency processing. Inspired by the encoder learned by \ac{convtasnet}, in addition to the logarithmically spaced filters, the proposed filterbank holds multiple gammatone filters at the same center frequency with varying phase shifts. We show that replacing the learned encoder with our proposed \ac{mpgtf} even leads to a \ac{sisnr} improvement of 0.7~dB. Furthermore, in contrast to using the learned encoder we show that the number of filters can be reduced from 512 to 128 without loss of performance.
\end{abstract}

\acresetall %

\section{Introduction}
\label{sec:intro}

The introduction of end-to-end systems to the problem of monaural speech separation has led to significant performance gains in recent years. In contrast, deep learning approaches such as Deep Clustering \cite{hershey_deep_2016}, \ac{PIT} \cite{yu_permutation_2017, kolbaek_multitalker_2017}, Deep Attractor Networks \cite{chen_deep_2017} and Chimera++ \cite{wang_alternative_2018} tackle the separation problem by transforming the input mixed speech into its \ac{STFT} representation and then training a network to find an optimal multiplicative mask for each speaker in this domain. These \ac{STFT} magnitude based approaches show a reasonable performance for the separation task but their performance is limited by the ideal mask calculation which typically does not include phase information.

These structural limitations are circumvented by end-to-end speech separation systems such as \acs{tasnet} \cite{luo_tasnet:_2018}, \acs{convtasnet} \cite{luo_conv-tasnet:_2019} or FurcaNext \cite{shi_end--end_2019, shi_deep_2019}. These systems introduce several changes to the \ac{STFT} magnitude based approaches: First, the training loss is defined in the time domain instead of the \ac{STFT} domain. Secondly, this time domain training loss allows for replacing the \ac{STFT} analysis and synthesis stages by learned convolutional analysis and synthesis filterbanks. As in \cite{luo_tasnet:_2018}, we refer to the convolutional analysis filterbank as the \textit{encoder} and the convolutional synthesis filterbank as the \textit{decoder}. Thirdly, the encoder and decoder exhibit a significantly lower frame size than the \ac{STFT} analysis-synthesis windows of state-of-the-art \ac{STFT}-based approaches \cite{hershey_deep_2016,yu_permutation_2017, kolbaek_multitalker_2017,chen_deep_2017,wang_alternative_2018}. These reduced frame sizes allow for a strongly reduced algorithmic latency if the separation network does not employ look-ahead.
And finally, at least for the \ac{convtasnet} and FurcaNext, the separation section of the network is implemented as a temporal convolutional network with a bottleneck structure \cite{lea_temporal_2017} instead of an architecture using \ac{LSTM} layers. In \cite{luo_conv-tasnet:_2019}, this modification has empirically shown to give better average results and to improve the robustness of these systems against time shifts of the input signal.

When replacing the deterministic \ac{STFT} analysis-synthesis structure by a learned encoder-decoder structure, the following general question arises: Should we use a well-understood, deterministic encoder (analysis filterbank) which is based on signal processing principles and possibly motivated by perceptual features? Or should we let the network run free and find a data-driven signal encoding for the given problem all by itself? This question has very recently gained attention and was investigated in several research papers such as \cite{heitkaemper_demystifying_2019}, \cite{von_neumann_end--end_2019} and \cite{pariente_filterbank_2019}. 

On a theoretical level, there are good arguments for both choices. Advocating for a learned encoder, we can argue that we might obtain a better network after training if the network has a high degree of freedom and given that we have diverse and comprehensive training data. Advocating for a deterministic, hand-crafted encoder such as an STFT analysis or a gammatone filterbank, we can argue that a smaller number of trainable parameters limits the variance of the model and its ability to overfit. Hand-crafting the feature extraction allows for a regularization of the optimization problem. While this may potentially lead to a worse training optimum, the network may potentially generalize better, in particular if the training data is limited.

In this work, we propose to replace the learned encoder of the network by a deterministic gammatone filterbank motivated by auditory features. By this, we want to enlighten the discussion on learned versus deterministic signal encoding for the speech separation problem. We will furthermore show that replacing the learned encoder with our proposed \ac{mpgtf} leads to an overall improvement from \SI{15.4}{\dB} to \SI{16.1}{\dB} measured in average \ac{sisnr} and allows to reduce the number of filters from 512 to 128 without performance loss.

We will first give an overview over our experimental framework in Section \ref{sec:framework} and then go into details about our proposed \ac{mpgtf} for the encoder in Section \ref{sec:filterbanks}. In Section \ref{sec:results} we present our results to then come to our conclusions in Section \ref{sec:conclusion}.

\section{Experimental Framework} \label{sec:framework}

\subsection{Conv-TasNet}
\begin{figure}%
 \centering
 \def\svgwidth{\linewidth}
\input{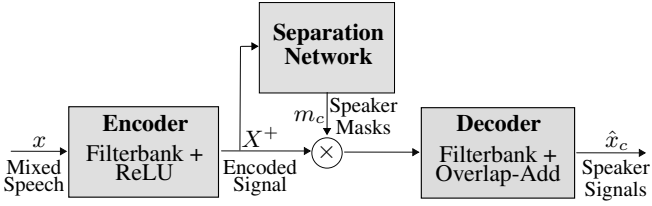}
 \caption{\ac{convtasnet} \cite{luo_conv-tasnet:_2019} architecture. In this work we experiment with the encoder and decoder stage while the separation network parameters remain untouched.}
 \label{fig:tasnet}
 \end{figure}

The architecture of the utilized neural network for speech separation is the \ac{convtasnet} architecture as presented in \cite{luo_conv-tasnet:_2019}. In Fig. \ref{fig:tasnet} we show the overall structure of \ac{convtasnet} which consists of three main structural elements, namely the encoder, the separation network and the decoder.

The \textit{encoder} can be characterized as a filterbank with $N$ filters of length $L$.
The output signal $X(n, i) \in \mathbb{R}^{N \times \lfloor T/D \rfloor} $ of the $n$-th filter $h_n^{\mathrm{Enc}}$ for the mixed speech input signal $x \in \mathbb {R}^T$ is calculated as 
\begin{equation} \label{eq:encfilterbank} 
    X(n, i) = \sum_{l=0}^{L-1} x(iD+l) h_n^{\mathrm{Enc}}(L-l)
\end{equation}
where $i$ is the frame index, $D$ is the frame shift and $T$ is the length of the input signal. %
As described in \cite{luo_tasnet:_2018, luo_conv-tasnet:_2019}, we provide the separation network a non-negative signal. Thus, we eliminate the negative values using
\begin{equation} \label{eq:relu}
X^+(n,i) = \max(0, X(n, i)).
\end{equation}
which can be implemented by a \ac{relu} layer. To account for this potential information loss, we propose to include phase inverted filters in the deterministic filterbank in Section~\ref{sec:mpgtf}.

The \textit{separation network} which is explained in detail in \cite{luo_conv-tasnet:_2019} 
 remains untouched for all our experiments with the hyper-parameters set fixed as shown in Table~\ref{tab:hyperparams}. We run the network in its non-causal configuration and use the global layer normalization method as detailed in \cite{luo_conv-tasnet:_2019}. At its output, the separation network provides a non-negative mask value $m_{c}(n,i) \in \mathbb{R}^{+}$ for the $c$-th speaker, all filters $n \in \{1,.., N\}$ and all time frames $i$. We obtain an estimate of each speaker in the encoded domain by the multiplication
\begin{equation}
    X_{c}^+(n,i) = X^+(n,i) m_{c}(n,i).
\end{equation}

In the \textit{decoder} we reconstruct the estimated time domain signal of the $c$-th speaker $\hat{x}_c \in \mathbb{R}^T$ by first calculating intermediary sums for each filter weight index $l$ and all time frames $i$ as 
\begin{equation} \label{eq:decoder} 
    x_c(l, i) = \sum_{n=0}^{N-1} X^+_c(n, i) h^{\mathrm{Dec}}_{N-n} (l)
\end{equation}
and then executing the overlap-add operation as
\begin{equation}
    \hat{x}_c(k) = \sum_{i=-\infty}^{\infty} x_c (k-iD ,i)
\end{equation}
where $h_n^{\mathrm{Dec}}$ holds the coefficients of the $n$-th filter of the decoder filterbank. In \cite{luo_conv-tasnet:_2019} the coefficients of $h_n^{\mathrm{Enc}}$ and $h_n^{\mathrm{Dec}}$ are initiliazed randomly and learned at training time. In this work we propose to replace the learned coefficients with the values of a deterministic filterbank. Also note that in \cite{luo_conv-tasnet:_2019} a matrix notation is used for description of the main structural elements of \ac{convtasnet} which simplifies the implementation of \ac{convtasnet} as a neural network. The notation as presented in this work is used to highlight signal processing properties of the encoder and decoder of \ac{convtasnet}. 

\begin{table}%
\begin{tabular}{|c|c|c|}
\hline
\textbf{Symbol} & \textbf{Description} & \textbf{Value} \\ \hline
N & Number of filters in encoder / decoder & \textit{varying} \\ \hline
L & Length of filters in samples & 16 \\ \hline
D & Frame shift in samples & 8 \\ \hline
B & \begin{tabular}[c]{@{}c@{}}Number of channels in bottleneck\\  and residual paths' $1 \times 1$-conv blocks\end{tabular} & 256 \\ \hline
H & \begin{tabular}[c]{@{}c@{}}Number of channels in\\ convolutional blocks\end{tabular} & 512 \\ \hline
P & Kernel size in convolutional blocks & 3 \\ \hline
X & \begin{tabular}[c]{@{}c@{}}Number of convolutional blocks\\  in each repeat\end{tabular} & 8 \\ \hline
R & Number of repeats & 4 \\ \hline
\end{tabular}
\caption{\ac{convtasnet} hyper-parameter settings. The parameter N is varied in our experiments and does only affect the encoder and the decoder of the network. All other parameters remain fixed for all carried out experiments.}
\label{tab:hyperparams}
\end{table}

\subsection{Network configurations}
For training and evaluating the network we use the WSJ0-MIX2 data set \cite{hershey_deep_2016} which is commonly used for the speaker-independent monaural speech separation task. It consists of mixtures of two speakers mixed at power ratios in between 0 and 10 dB. The total amount of audio is roughly 20 hours in the training set, 5 hours in the validation set and 3 hours in the test set. For measuring the test set performance for a certain network configuration, we consider the training epoch where the validation set loss is minimal.  The network training procedure is stopped when the validation set error has not improved within the last 10 epochs. All our experiments and evaluations are carried out at \SI{8}{\kHz} sampling rate.

\subsection{Evaluation metrics} 

For performance comparison of all tested configurations and for use as a training objective we use the \ac{sisnr} \cite{luo_tasnet:_2018} of a speech signal calculated as

\begin{equation} 
    s_{\mathrm{target}}:= \frac{\langle\hat{s},s\rangle s}{||s||^2} ,
\end{equation}
\begin{equation} 
    e_{\mathrm{noise}}:= \hat{s} - s_{\mathrm{target}} ,
\end{equation}
\begin{equation} \label{eq:sisnr}
    \mathop{\mbox{$\mathrm{SI}$-$\mathit{\mathrm{SNR}}$}} := 10\log_{10} \frac{||s_{\mathrm{target}}||^2}{||e_{\mathrm{noise}}||^2}
\end{equation}
where  $s, \hat{s} \in \mathbb{R}^T$ denote the clean and the estimated speech signals, $\langle\cdot, \cdot\rangle$ denotes the scalar product and $||\cdot||^2$ denotes the signal power. For the reported test set results we average the \ac{sisnr} improvements over all 3000 mixtures.

\section{Filterbank configurations} \label{sec:filterbanks}

\begin{figure}
\begin{tikzpicture}

\begin{groupplot}[group style={group size=5 by 1, horizontal sep=0.5cm}]
\nextgroupplot[
tick align=outside,
tick pos=left,
title={Coeffs.},
x grid style={white!69.01960784313725!black},
xlabel={Time (ms)},
xmin=-0.5, xmax=15.5,
xtick style={color=black},
xtick={0,15},
xticklabels={0,2},
y grid style={white!69.01960784313725!black},
ylabel={Filter index $n$},
y label style={at={(-0.1,0.5)}},
ymin=-0.5, ymax=127.5,
ytick style={color=black},
ytick={0,127},
yticklabels={1,128},
height=0.7\linewidth,
width=0.32\linewidth
]
\addplot graphics [includegraphics cmd=\pgfimage,xmin=-0.5, xmax=15.5, ymin=-0.5, ymax=127.5] {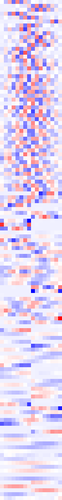};

\nextgroupplot[
tick align=outside,
tick pos=left,
title={$|\mathrm{FFT}|$},
x grid style={white!69.01960784313725!black},
xlabel={Freq. (Hz)},
xmin=-0.5, xmax=8.5,
xtick style={color=black},
xtick={0,8},
xticklabels={0,4000},
ytick style={draw=none},
yticklabels={},
ymin=-0.5, ymax=127.5,
height=0.7\linewidth,
width=0.32\linewidth
]
\addplot graphics [includegraphics cmd=\pgfimage,xmin=-0.5, xmax=8.5, ymin=-0.5, ymax=127.5] {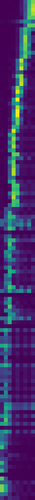};

\nextgroupplot[
xshift=1cm,
tick align=outside,
tick pos=left,
title={Coeffs.},
x grid style={white!69.01960784313725!black},
xlabel={Time (ms)},
xmin=-0.5, xmax=15.5,
xtick style={color=black},
xtick={0,15},
xticklabels={0,2},
y grid style={white!69.01960784313725!black},
ylabel={Filter index $n$},
y label style={at={(-0.1,0.5)}},
ymin=-0.5, ymax=127.5,
ytick style={color=black},
ytick={0,127},
yticklabels={1,128},
height=0.7\linewidth,
width=0.32\linewidth
]
\addplot graphics [includegraphics cmd=\pgfimage,xmin=-0.5, xmax=15.5, ymin=-0.5, ymax=127.5] {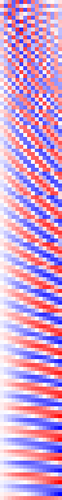};

\nextgroupplot[
tick align=outside,
tick pos=left,
title={$|\mathrm{FFT}|$},
x grid style={white!69.01960784313725!black},
xlabel={Freq. (Hz)},
xmin=-0.5, xmax=8.5,
xtick style={color=black},
xtick={0,8},
xticklabels={0,4000},
ytick style={draw=none},
yticklabels={},
ymin=-0.5, ymax=127.5,
height=0.7\linewidth,
width=0.32\linewidth
]
\addplot graphics [includegraphics cmd=\pgfimage,xmin=-0.5, xmax=8.5, ymin=-0.5, ymax=127.5] {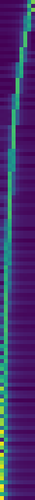};
\end{groupplot}

\node at ({$(current bounding box.south west)!0.05!(current bounding box.south east)$}|-{$(current bounding box.south west)!-0.06!(current bounding box.north west)$})[
  scale=1.1,
  anchor=base west,
  text=black,
  rotate=0.0
]{(a) Learned Filterbank \ \ \ \ \ \ \ \ \ \ \ \ \ (b) Proposed MP-GTF     };

\end{tikzpicture}
\caption{(a) Time and frequency domain representation of all filters  $h_n^{\mathrm{Enc}}$ for the encoder filterbank learned by \ac{convtasnet} with random initialization. (b) Time and frequency domain representation for all filters $h_n^{\mathrm{Enc}}$ of the proposed \ac{mpgtf}.  The learned filterbank was sorted by the filter's peak values in the frequency domain and the number of filters is $N=128$.}
\label{fig:encoderPlots}
\end{figure}

The goal of this work is to investigate if the learned filterbanks of the end-to-end system \ac{convtasnet} is the key to its success, or if using a well defined deterministic filterbank performs equally well or even better. We have decided to choose modified auditory gammatone filterbanks for the deterministic filterbank for multiple reasons. Most importantly, an \ac{agtf} resembles the signal encoding in human auditory perception. As speech is a central sound source for humans, we assume that the human auditory system provides a signal representation which facilitates separation of different speakers. A second reason for using auditory gammatone filterbanks is their non-linear spacing of center-frequencies, a structure that we also observe in the filterbanks learned by \ac{convtasnet}. In Figure~\ref{fig:encoderPlots} we plot the time domain and frequency domain representations of both the learned encoder filterbank and as well as for our proposed \ac{mpgtf} presented in Section~\ref{sec:mpgtf}. In the frequency representation of the learned filterbank, we can observe more filters that focus on the lower frequency regions than filters focusing on higher regions. We now first present the construction of a common \ac{agtf} as presented in \cite{patterson_efficient_1988, glasberg_derivation_1990} and then describe our modifications for usage in \ac{convtasnet}.

\subsection{Auditory gammatone filterbank (A-GTF)}

An \ac{agtf} resembles the patterns of basilar membrane motion in the human auditory system. The filterbank consists of non-linearly spaced narrow-band filters $\gamma_{f_c}$ with an increasing bandwidth over the filter's center frequency $f_c$. The impulse response $\gamma_{f_c} (t) \in \mathbb{R}$ of a real-valued gammatone filter is given by a gamma probability distribution function multiplied by a sinusoidal tone according to \cite{patterson_efficient_1988} as
\begin{equation} \label{eq:gammatoneIR}
    \gamma (t) = a \, t^{(p-1)} e^{-2\pi b t}  \cos{(2\pi f_c t + \phi)}
\end{equation}
where $f_c$ denotes the center frequency, $\phi$ the phase shift, $a$ the amplitude, $t>0$ the time in seconds, $p$ the filter order and $b$ the filter bandwidth parameter. \cite{patterson_complex_1992} states that the filter order $p$ measured for the human auditory system is in between 3 and 5.

The filter bandwidth parameter $b$ and the center frequencies $f_c$ are commonly determined using the concept of the \ac{erb}. The \ac{erb} estimates the bandwidth of the filters in the human auditory system using the simplification that each filter is a rectangular band-pass filter. The \ac{erb} has been found empirically as the following function over the center frequency~$f_c$ \cite{hohmann_frequency_2002}:
\begin{equation} \label{erb}
    \mathrm{ERB}(f_c) = 24.7 + \frac{f_c}{9.265}.
\end{equation}

The center frequencies of the human auditory filterbank are placed equally distant on the so called \ac{erb} scale which is derived by integrating $1/\mathrm{ERB}(f_c)$ across frequency \cite{glasberg_derivation_1990} resulting in 
\begin{equation} \label{nextErb}
    \mathrm{ERB}_{\mathrm{scale}}(f_{\SI{}{\Hz}}) = 9.265 \log ( 1+ \frac{f_{\SI{}{\Hz}}}{24.7 \times 9.265}).
\end{equation}
where $f_{\SI{}{\Hz}}$ denotes the frequency in Hertz. Thus, the relation between the acoustic frequency $f_{\SI{}{\Hz}}$ and the \ac{erb} scale is given by a nonlinear logarithmic equation.
A neighboring frequency with a distance of 1 on the \ac{erb} scale is then calculated as
\begin{equation} \label{eq:nextErb}
    f_{\mathrm{next}} = \mathrm{ERB}_{\mathrm{scale}}^{-1}(\mathrm{ERB}_{\mathrm{scale}}(f_0) +1)
\end{equation}
where $\mathrm{ERB}_{\mathrm{scale}}^{-1} (f_{\mathrm{ERB}})$ is the inverse of $\mathrm{ERB}_{\mathrm{scale}}(f_{\SI{}{\Hz}})$ with $f_{\mathrm{ERB}}$ denoting a frequency represented on the \ac{erb} scale.

To construct a complete \ac{agtf} we typically define a lowest center frequency at around \SI{50}{\Hz} and define an upper limit for the filter center frequencies at around \SI{8000}{\Hz}. By using (\ref{eq:nextErb}), we obtain all center frequencies of the filterbank by starting with the lowest center frequency and then iteratively going to higher frequencies until we reach the specified upper limit. For each center frequency we generate the time domain filters for the \ac{agtf} via the impulse response function given in (\ref{eq:gammatoneIR}). As detailed in \cite{hohmann_frequency_2002} the filters are normalized by their peak value in the freqency domain by varying the amplitude parameter $a$ in (\ref{eq:gammatoneIR}) and the phase is set as $\phi=0$.

\subsection{Multi-phase gammatone filterbank (MP-GTF)} \label{sec:mpgtf}
To utilize a gammatone filterbank as a deterministic encoder of \ac{convtasnet} as described in Section \ref{sec:framework}, we propose several adaptations to the  \ac{agtf}.  We name our proposed analysis filterbank for \ac{convtasnet} the \textit{\acf{mpgtf}}\protect\footnote{We provide Python code for construction of our proposed \ac{mpgtf} under \url{https://github.com/sp-uhh/mp-gtf} }.

First, to keep the implementation close to original \ac{convtasnet}, we truncate the infinite impulse response (\ref{eq:gammatoneIR}) to a short length of \SI{2}{\ms} which corresponds to a length of 16~samples at \SI{8}{\kHz} sampling rate. This choice of a short filter length allows to keep the system latency low when using a causal separation network. Setting the filter order in between 3 and 5 as in the \ac{agtf} leads to filters at low center frequencies which do not contain the peak amplitude of the impulse response (\ref{eq:gammatoneIR}). For this reason, in our modified \ac{mpgtf} we use the filter order of $p=2$, which shifts the amplitude peaks towards the time origin and inside the incorporated time span of the gammatone. For this filter order, the bandwidth parameter for the gammatone at the center frequency $f_c$ equates to $b=ERB(f_c)/1.57$ according to \cite[Eq. 14]{hohmann_frequency_2002}.

Our second modification of the \ac{agtf} for use in \ac{convtasnet} is to reduce the frequency range of the filter bank. The lowest center frequency of our proposed filterbank is chosen as \SI{100}{\Hz} while the upper range limit is set to \SI{4000}{\Hz} which is the Nyquist frequency for the utilized sampling rate of \SI{8000}{\Hz}. With the same construction method as for the \ac{agtf} and the same spacing of 1 on the \ac{erb} scale, 24 center frequencies can be fitted within this limited frequency range.

As the third adaption and to meet the non-negativity constraint for the input to the separation network of \ac{convtasnet}, for each filter $h_n^{\mathrm{Enc}}$ in the encoder, we include the negative version $-h_n^{\mathrm{Enc}}$ in our filterbank. Thus, we ensure that for each center frequency and each frame there is at least one filter that contains energy, if the original signal holds energy at this particular frequency. For a gammatone filter as in (\ref{eq:gammatoneIR}) we can obtain this negative filter with a phase shift of $\pi$ as $\phi^{\mathrm{inv}} = \phi + \pi$. 

As the last adaption, we propose a method to control the number of filters $N$ in the encoder filterbank by introducing multiple filters with different phase shifts for the same center frequency. The minimum number of $N$ based on our proposed construction is 48 due to the 24 center frequencies and the need to include the phase inverted version of each filter. If $N>48$, we introduce multiple filters at each center frequency by choosing multiple values of $\phi$. In total, we can only choose $N/2$ filters freely, as the other half is determined by the need for the phase inverted filters. This also implies that, based on our construction, $N$ must be even. For a fixed center frequency $f_c$ we introduce a total number of $\lfloor N/2/24 \rfloor$ phases where 24 is the number of center frequencies. As the division might hold a remainder, we distribute the remaining number of phase shifts to the lowest center frequencies. Once the number of phase shifts for a specific frequency is determined, we place the phase shifts equidistantly on the interval $[0,\pi)$ with the first phase shift being $0$. We construct the first half of the filters at this center frequency by inserting these different values of $\phi$ into the gammatone impulse response (\ref{eq:gammatoneIR}). The other half of filters for this center frequency is generated by inserting all inverted phase shifts in the interval $[\pi, 2\pi)$ into the gammatone impulse response (\ref{eq:gammatoneIR}). In Figure \ref{fig:encoderPlots}, we plot the time and frequency domain representation of our proposed \ac{mpgtf} with $N=128$.

\section{Results} \label{sec:results}

In Table~\ref{tab:resultsByConfig}, we present the results of our experiments for different configurations of the encoder and decoder of \ac{convtasnet} with the number of filters fixed at $N=512$. For the original configuration as in \cite{luo_conv-tasnet:_2019} we report a test set performance of \SI{15.4}{\dB} where \cite{luo_conv-tasnet:_2019} reports a comparable performance of \SI{15.6}{\dB}. When replacing the encoder with our proposed deterministic \ac{mpgtf} and the decoder with the pseudo-inverse \cite{penrose_generalized_1955} of the \ac{mpgtf}, we reach a comparable performance of  \SI{15.4}{\dB}. Table~\ref{tab:resultsByConfig} also shows that the gap between the training and the test set is reduced from 3.9~dB to 2.7~dB in this configuration which suggests that a deterministic encoder and decoder combination is less prone to over-fitting. The optimal choice in terms of overall performance is to use our proposed \ac{mpgtf} in the encoder and to use a learned decoder which results in a performance of \SI{15.9}{\dB} for $N=512$. The decoder weights are initialized with the pseudo-inverse of the \ac{mpgtf} in this setting. A small number of reruns of all configurations has shown little variance for the test set performances.

\begin{table}%
\centering
\begin{tabular}{c|c|c|c|l}
\textbf{Encoder} & \textbf{Decoder} & \textbf{N} & \multicolumn{2}{c}{\textbf{SI-SNRi (dB)}} \\ \hline

\multicolumn{1}{l}{} & \multicolumn{1}{l}{} & \multicolumn{1}{l}{} & \multicolumn{1}{l|}{\textbf{Train}} & \textbf{Test} \\ \hline  \hline
Learned & Learned & 512 & 19.3 & 15.4 \\ \hline
\ac{mpgtf} & Learned & 512 & 19.0 & \textbf{15.9} \\ \hline
\ac{mpgtf} & \ac{mpgtf} Pseudo Inv.  & 512 & 18.1 & 15.4 \\ 
\end{tabular}
\caption{\ac{sisnr} improvements on WSJ0-MIX2 training and test set for different configurations of the encoder and decoder of \ac{convtasnet} for $N=512$ filters. Higher is better.}
\label{tab:resultsByConfig}
\end{table}

In Table~\ref{tab:resultsByN}, we present the average test set performance for a varying number of filters $N$. Similar to \cite{luo_conv-tasnet:_2019}, we found that for the original \ac{convtasnet} with learned filterbanks, the optimal performance is reached for $N=512$ while lowering the number of filters to $N=128$ slightly decreases the performance by a value of \SI{0.2}{\dB}. In contrast, when replacing the learned encoder with our proposed deterministic \ac{mpgtf} we find that we can lower the number of filters to $N=128$ without performance loss, even resulting in the overall best model with an average \ac{sisnr} performance of \SI{16.1}{\dB}. %
When setting $N=48$, i.e. without redundant phase information in the encoder, the performance drops to \SI{14.4}{\dB}. This supports the hypothesis that amending the filters of the encoder with phase shifted versions is important for the performance gain of \ac{convtasnet}.

\begin{table}%
\centering
\begin{tabular}{c|c|c|c}
\textbf{Encoder} & \textbf{Decoder} & \textbf{N} & \textbf{SI-SNRi (dB)} \\ \hline
\hline
Learned & Learned &\textbf{512} & \textbf{15.4} \\ \hline
Learned & Learned &128 & 15.2 \\ \hline
\ac{mpgtf} & Learned &512 & 15.9 \\ \hline
\ac{mpgtf} & Learned &\textbf{128} & \textbf{16.1} \\ \hline
\ac{mpgtf} & Learned & 64 & 15.4 \\ \hline
\ac{mpgtf} & Learned & 48 & 14.4 \\
\end{tabular}
\caption{\ac{sisnr} improvements on WSJ0-MIX2 test set for different values of the number of filters $N$ with a learned decoder. %
}
\label{tab:resultsByN}
\end{table}

In addition to the values presented in the tables, we carried out performance analyses based on the difference of the median fundamental frequencies of the speakers within a mixture. As we have shown in \cite{ditter_influence_2019}, the median fundamental frequency difference is an important influencing factor to the performance of monaural speech separation systems and it is of importance to improve the performance especially for mixtures of similar speakers where the fundamental frequency difference is below \SI{50}{\Hz}. For our best model we were able to improve the performance in this important region of fundamental frequency difference by \SI{0.9}{\dB} in contrast to \ac{convtasnet} with a learned filterbank. 

Our findings imply that for the training data and architecture used for \ac{convtasnet}, the learned encoder is not critical for the performance gain of \ac{convtasnet} with respect to competing approaches. In contrast, a hand-crafted deterministic encoder filterbank performs even slightly better. In particular, it may reduce the gap between training and testing performance and also gives better results for difficult speaker pairs with similar fundamental frequencies, both of which indicates an increased robustness.  Also, compared to the learned encoder, the number of filters $N$ can be reduced without sacrificing performance.

\section{Conclusions and Future Work} \label{sec:conclusion}

In this paper, we analyzed if the learned encoder is critical for the recent success of \ac{convtasnet} for monaural speech separation.
We have shown that the overall system performance does not decrease when the learned encoder is replaced by our proposed deterministic \acf{mpgtf}. Instead, we can even measure a slight performance increase of \SI{0.7}{\dB} in average \ac{sisnr}. Furthermore with \ac{mpgtf} as the encoder filterbank, the number of filters can be reduced from 512 to 128 filters without compromising the overall performance. With the encoder filterbank set to the proposed \ac{mpgtf} and the decoder set as its pseudo-inverse, we showed that the overall performance is similar to the fully end-to-end system while over-fitting on the training data is reduced. 
In future work, it should be investigated if our findings hold for a broader range of speech processing tasks that can be tackled by neural networks such as monaural speech enhancement. Furthermore, we suggest to investigate if our proposed encoder leads to a more robust system by testing it with a larger variety of test data than given by the WSJ0-MIX data set. As a deterministic approach seems less prone to over-fitting, the proposed \ac{mpgtf} may provide for a more robust system in real-world applications.

\atColsBreak{\vskip5pt}

\bibliographystyle{IEEEbib}
\bibliography{refs}

\end{document}